\documentclass[12pt]{article}
\pdfoutput=1
\usepackage{authblk}
\usepackage{bm}
\usepackage{mathrsfs}
\usepackage{amsbsy}
\usepackage{graphicx}
\usepackage{subfigure}
\usepackage{amsmath}
\usepackage{amssymb}
\usepackage{braket}
\usepackage[numbers,sort&compress]{natbib}
\usepackage{lipsum}
\usepackage[]{hyperref}
\newcommand\blfootnote[1]{
\begingroup
\renewcommand\thefootnote{}\footnote{#1}
\addtocounter{footnote}{-1}
\endgroup
}

\newcommand{\mpref}[1]{Figure.\ref{#1}}
\numberwithin{equation}{section}

\begin{document}
	
	\begin{center}
		{\bf Entanglement Islands in Generalized Two-dimensional Dilaton Black Holes}\\

		\vspace{1.6cm}
		{\textbf{Ming-Hui Yu}$^{1}$, ~\textbf{Xian-Hui Ge}$^{1,2}$}\blfootnote{* Corresponding author. gexh@shu.edu.cn}
		\vspace{0.8cm}
		
		$^1${\it Department of Physics, Shanghai University, Shanghai 200444, China} \\

		\vspace{1.6cm}

		\begin{abstract}
		The Fabbri--Russo model is a generalized model of a two-dimensional dilaton gravity theory with various parameters ``$n$'' describing various specific gravities. Particularly, the Russo--Susskind--Thorlacius gravity model fits the case $n=1$. In the Fabbri--Russo model, we investigate Page curves and the entanglement island. Islands are considered in eternal and evaporating black holes. Surprisingly, in any black hole, the emergence of islands causes the rise of the entanglement entropy of the radiation to decelerate after the Page time, satisfying the principle of unitarity. For eternal black holes, the fine-grained entropy reaches a saturation value that is twice the Bekenstein--Hawking entropy. For evaporating black holes, the fine-grained entropy finally reaches zero. The parameter ``$n$'' significantly impacts the Page curve at extremely early times. However, at late times and large distance limit, the impact of the parameter ``$n$'' is a subleading term and is exponentially suppressed. As a result, the shape of Page curves is ``$n$''-independent in the leading order. Furthermore, we discuss the relationship between islands and firewalls. We show that the island is a better candidate than firewalls for encountering the quantum entanglement-monogamy problem. Finally, we briefly review the gravity/ensemble duality as a potential resolution to the state conundrum resulting from the island formula.

		\end{abstract}
	\end{center}
\newpage
\tableofcontents
\newpage

\section{Introduction} \label{Introduction}
\qquad One of the most important and urgent objectives of modern physics is studying the theory of quantum gravity. Black holes serve as fantastic study tools for this theory. When the quantum field theory (QFT) is introduced into curved spacetime, black holes can emit Hawking radiation \cite{HR}. However, a series of issues also follow. One of the most famous issues is the black hole information loss paradox, or the information paradox for short, which was proposed by Stephen Hawking in 1976 \cite{Paradox}. The entanglement entropy of the radiation increases with time and is divergent at late times due to the continuous occurrence of the Hawking radiation. This is inconsistent with the unitary evolution of quantum mechanics (QM). Therefore, the information paradox challenges QM, general relativity (GR), thermodynamics, and many other fundamental fields of modern physics. A specific solution to this issue is to obtain the Page curve directly without the unitary evaporation hypothesis during the evaporation of a black hole \cite{PC1, PC2}. Undoubtedly, the discovery of the Anti-de Sitter (AdS)/conformal field theory (CFT) duality opens up a wider field of investigation for the information paradox \cite{adscft}. Based on the view that the evolution of the pure high-energy state in the CFT is unitary, we insist that the AdS/CFT duality is right, implying the evolution of black holes in pure states in AdS spacetime is also unitary. Thus, the general consensus is that information is conserved for AdS black holes, although the precise specifics and methods are yet unknown. Moreover, quantum entanglement provides a starting point for addressing this issue. The Ryu-Takayanagi (RT) formula for the holographic entanglement entropy is not only a valid demonstration of the AdS/CFT dual conjecture, but also paves the way for the microscopic origin of the black hole entropy \cite{RT}. After considering the von-Neumann entropy contributed by CFT at the boundary of the bulk spacetime, the RT formula is extended to the quantum RT formula (QRT) \cite{QRT} and the quantum extremal surface (QES) prescription with high-order quantum corrections \cite{QES}. Surprisingly, one obtains the opposite result of Hawking's calculation when using the QES prescription to calculate the entanglement entropy for AdS black holes \cite{bulk entropy}. At late times, the (quantum extremal) island locates near the event horizon. This construction renders the degree of freedom (DOF) inside the island belonging to the DOF of outside radiation. Therefore, the entanglement entropy decreases rather than increases at late times, which reproduces the unitary Page curve \cite{island rule, entanglement wedge}. Physicists summarize this approach as the island paradigm \cite{island rule, review}. The ansatz for calculating the fine-grained entropy is called the island formula \cite{review}
\begin{equation}
S_{\text{rad}}=\text{Min} \Bigg \{ \text{Ext} \bigg [ \frac{\text{Area}(\partial I)}{4G_N} +S_{\text{matter}} (I \cup R) \bigg] \Bigg \}.   \label{island formula}
\end{equation}
In the above equation, the LHS is the fine-grained entropy of the radiation; on the RHS, the terms inside the square bracket are called the generalized entropy of radiation, where $\partial I$ represents the boundary of the island and $I \cup R$ is the union of the island and the radiation. This equation instructs us to first evaluate the generalized entropy before extremizing it. The smallest value among candidates is the correct answer.\\
\indent Besides, the island formula can also be derived equivalently from the gravitational path integral \cite{replica1,replica2}. The $n$-th order Renyi entropy is obtained using the replica trick
\begin{equation}
S_A ^{(n)}=\frac{1}{1-n} \log \bigg[ \text{Tr} \big(\rho_A^{(n)}\big) \bigg], \label{renyi}
\end{equation}
where $\rho_A$ is the reduced density matrix. The von-Neumann (fine-grained) entropy is given by
\begin{equation}
S_{\text{von}}=\lim_{n\to 1}S_A^{(n)} = -\text{Tr} (\rho_A \log \rho_A). \label{von entropy}
\end{equation}
Along with the original Hawking saddle, we also need to consider the replica wormholes saddle provided by the island while performing the explicit computation. The unitary result is produced at late times when the replica wormholes dominate.\\
\indent The Page curve is first obtained by evaporating black holes in Jackiw--Teitelboim (JT) gravity \cite{island rule, JT1, JT2}. Later work extends it to eternal black holes \cite{eternal bh}. Further, there are many interesting developments, such as some other two-dimensional (2D) asymptotically flat or AdS black holes \cite{2d1,2d2,2d3,2d4,2d5,2d6,2d7,2d8,2d9}, higher-dimensional black holes \cite{high1,high2,high3,high4,high5,high6,high7,high8,high9,high10,high11,high12,high13,high14,high15,high16,high17,high18,high19,high20,high21,high22,high23,high24,high25,high26,high27,high28}, charged black holes \cite{charge1,charge2,charge3,charge4,charge5,charge6,charge7}, and some models of the universe \cite{ds1,ds2,ds3,ds4,ds5,ds6}. Interestingly, there also some work on complexity related to entanglement islands \cite{complexity1,complexity2,complexity3}. Here is an incomplete list because the field is always developing. Note that most works are performed in the background of eternal black holes. Although we can more easily understand the information paradox in eternal black holes, the actual evaporating black hole is the key to obtaining Page curves. Therefore, the Page curve for 2D evaporating black holes is the major focus of this paper.\\
\indent A contender for the quantum gravity theory, the 2D gravity theory is crucial for studying black holes. On the one hand, 2D models are widely studied as they are symmetric and simpler to calculate analytically. On the other hand, they come from the four-dimensional theory through the dimensional reduction, and we hope that the keen insight from these models can be applied to higher-dimensional problems. One of the most famous models---the Callan--Giddings--Harvey--Strominger (CGHS) model \cite{cghs} or the Russo--Susskind--Thorlacius (RST) model with considerations for back-reaction effects \cite{rst}---has the advantage of being analytically solvable and offers a semiclassical theory of the back-reaction of the Hawking radiation. It can describe the entire process of a black hole forming and evaporating (including the back-reaction) in classical physics. This paper focuses on the generalized 2D theory of gravity---the Fabbri-Russo (FR) model \cite{fr1,fr2}. This model represents a one-parameter family of exactly solvable models and includes the RST model as the special case. We use the island paradigm to reproduce the Page curve of this model and provide some more general conclusions for 2D models. We also discuss the relationship between firewalls and islands from the perspective of entanglement monogamy. We show that the island is a better candidate than the firewall. Finally, we review the latest report on the state paradox and the gravity/ensemble duality.\\
\indent This paper is organized as follows. In Sec.\ref{Exactly Solvable Model}, we introduce the FR model and the entanglement entropy calculation in CFT. The solutions of vacuum and black holes are mentioned. In Sec.\ref{Evaporating Black Holes}, we calculate the entanglement entropy of the radiation for evaporating black holes by the island formula \eqref{island formula}. The results show that islands emerged at late times and led to a unitary Page curve. In Sec.\ref{Eternal Black Holes}, the same procedure is applied to the eternal black hole. We obtain similar results for the evaporating black hole. In Sec.\ref{Scrambling Time}, we demonstrate the corresponding Page curves and the scrambling time based on the previous results. We also investigate the effect of the parameter $n$ on these results. In Sec.\ref{Firewalls}, we discuss the relations between the island and the firewall. Furthermore, based on the latest report, we briefly review the paradox related to the state of the Hawking radiation. The final discussion and conclusion are presented in Sec.\ref{Conclusion}. The Planck units are applied $\hbar=c=k_B=1$ throughout.

\section{Exactly Solvable 2D FR Model} \label{Exactly Solvable Model}
\qquad We are inspired by the special 2D RST action to obtain the generalized 2D gravity \cite{rst}, we consider the following action \cite{fr1}:
\begin{subequations}
\begin{align}
S_\text{cl}&=S_0+S_{\text{CFT}}, \label{full action} \\
S_0&=\frac{1}{2\pi} \int d^2x \sqrt{-g} \bigg \{ e^{-\frac{2\phi}{n}} \bigg[R+\frac{4}{n} (\nabla \phi)^2\bigg]+4\lambda^2  e^{-2\phi}\bigg \},  \label{fr action} \\
S_{\text{CFT}}&=-\frac{1}{4\pi} \sum_{i=1}^{N} \int d^2x \sqrt{-g} (\nabla f_i)^2,   \label{cft action}
\end{align}
\end{subequations}
where $\phi$ is a dilaton field, $\lambda$ is the cosmological constant, $f_i$ is a set of $N$ massless scalar fields, and $n$ is a parameter characterizing the different theory. Similar to the RST model \cite{rst}, we choose the conformal gauge:
\begin{equation}
g_{\pm \pm}=0,  \qquad g_{\pm \mp}=-\frac{1}{2} e^{2 \rho}.   \label{conformal gauge}
\end{equation}
The metric in this gauge is given by
\begin{equation}
ds^2=-e^{2 \rho} dx^+dx^-,  \label{2d metric}
\end{equation}
with the light-cone coordinate $x^{\pm}=x^0 \pm x^1$.\\
\indent After a field redefinition, the action \eqref{full action} can be written in the ``free field'' form \cite{fr1}
\begin{equation}
S_{\text{cl}}=\frac{1}{\pi} \int d^2 x  \bigg[ \frac{1}{\kappa} \bigg( -\partial_+ \chi \partial_- \chi + \partial_+ \Omega \partial_- \Omega \bigg) +\lambda^2  \label{free field} e^{2(\chi-\Omega)/ \kappa} +\frac{1}{2} \sum_{i=0}^{N} \partial_+ f_i \partial_- f_i \bigg],
\end{equation}
where
\begin{subequations}
\begin{align}
\kappa &=\frac{N}{12},  \label{kappa} \\
\chi &=\kappa \rho +e^{-\frac{2\phi}{n}}+\bigg(\frac{1}{2n}-1 \bigg) \kappa \phi, \label{chi} \\
\Omega &= e^{-2\phi/n} +\frac{\kappa}{2n} \phi,  \label{conformal factor1}
\end{align}
\end{subequations}
where $n$ is a real positive number. The special case $n=1$ corresponds to the RST model \cite{rst}. For simplicity, in this paper, we only consider the parameter range\footnote{For $n<0$, the geometry does not exist. When $n>2$, the corresponding geometry is very different, where the singularity is null. One can refer to \cite{fr2} for details.} $0<n<2$, while for the case of critical values $n=0$ and $n=2$, we discuss them in the Appendix \ref{appendix}.\\
\indent One should also note that for real $\phi$, $\Omega$ has a lower value
\begin{equation}
\Omega \ge \Omega_{\text{crit}}= \frac{\kappa}{4}-\frac{\kappa}{4} \log \frac{\kappa}{4}. \label{critical}
\end{equation}
Here, we only give a qualitative physical interpretation. As the 2D gravity theory results from the dimensional reduction of the 4D theory, $\Omega$ corresponds to the area of the transverse two-sphere surface. The condition \eqref{critical} allows us to patch the vacuum at the endpoint of evaporation. Therefore, the curve \eqref{critical} is converted from a time-like curve to a space-like curve, i.e., this critical curve can be regarded as the boundary of spacetime, which is similar to the curve $r=0$ in the spherically symmetric reduction of 4D Minkowski spacetime. The explicit calculation is given in the Appendix \ref{appendix}.\\
\indent The equations of motion (EOM) derived by \eqref{free field} are
\begin{equation}
\partial_+\partial_-(\chi-\Omega)=0, \qquad \partial_+\partial_- \chi =-\lambda^2 e^{2(\chi-\Omega)/\kappa}.  \label{eom1}
\end{equation}
We choose the ``Kruskal'' gauge, where
\begin{equation}
\chi = \Omega.   \label{kruskal gauge1}
\end{equation}
Then, it implies that
\begin{equation}
\phi = \rho_K,  \label{kruskal gauge2}
\end{equation}
and
\begin{equation}
\Omega=e^{- 2 \rho_K /n}+\frac{\kappa}{2n} \rho_K,  \label{conformal factor2}
\end{equation}
where $\rho$ is denoted as $\rho_K$ in the Kruskal coordinate. In this gauge, the most general static solution for \eqref{eom1} is
\begin{equation}
\Omega=\chi= -\lambda^2 x^+x^- +Q \log (\lambda^2 x^+x^-) +\frac{M}{\lambda},  \qquad Q,M= \text{constant}. \label{general solution1}
\end{equation}
Afterward, we set $\lambda \equiv 1$ for convenience.

\subsection{Vacuum Solution}
\qquad The simplest case of the above solution is the vacuum, where the Ricci scalar $R$ and the integrate constant $M$ vanish. Accordingly, the conformal factor is\footnote{For convenience, we omit the subscript ``K'' for Kruskal coordinate hereafter.}
\begin{subequations}
\begin{align}
e^{-2\rho/n}&=e^{-2\phi/n}=-x^+x^-, \label{vacuum} \\
\Omega_{\text{vac}}&=-x^+x^--\frac{\kappa}{4} \log(-x^+x^-). \label{vacuum conformal factor}
\end{align}
\end{subequations}
 The most general theory that differs from and incorporates the RST model may be found. Referring to the RST model \cite{rst}, the general theory obeys the following requirements. The action should contain an RST-like term and a conformal anomaly term to describe the back-reaction effect, resulting in the effective action for the large $N$-limit, as indicated by the sum \cite{fr1}
\begin{equation}
\begin{split}
S_{\text{eff}}&=S_{\text{cl}}+S_{\text{RST}}+S_{\text{anom}} \\
&=\frac{1}{2\pi} \int d^2 x \sqrt{-g}  \bigg[ e^{-2 \phi/n} \bigg(R+\frac{4}{n}(\nabla \phi)^2\bigg) +4e^{-2\phi} -\frac{1}{2} \sum_{i=1}^N (\nabla f_i)^2 \\
&+\kappa \bigg(\frac{1-2n}{2n} \phi R +\frac{n-1}{n} (\nabla \phi)^2 -\frac{1}{4}  R \Box^{-1} R  \bigg) \bigg].  \label{effective action}
\end{split}
\end{equation}
In the last line, the first two terms are one-loop quantum correction terms similar to the RST model $(n=1)$ and the last term is a non-local Polyakov term, where the symbol $\Box^{-1}$ is the scalar Green function. In the conformal gauge \eqref{conformal gauge}, it has a simple form, where $\Box^{-1}R=2\rho$.\\
\indent We first consider the classical limit, in which $\hbar$ is turned to zero. By recovering the effective action \eqref{effective action} with $\hbar$, the last three terms vanish. Then, we obtain the action \eqref{fr action}. The EOM in the conformal gauge \eqref{conformal gauge} is derived by variation of the metric
\begin{subequations}
\begin{align}
&-\frac{4}{n} \partial_+ \phi \partial_- \phi +\frac{2}{n} \partial_+ \phi \partial_- \phi -e^{\frac{2-2n}{n}\phi+2\rho}=0,  \label{variation1} \\
& \frac{n}{2} \partial_+ \phi \partial_- \phi +\frac{4}{n^2} \partial_+ \phi \partial_- \phi -\frac{4}{n}\partial_+ \phi \partial_- \phi +e^{\frac{2-2n}{n}\phi+2\rho}=0, \label{variation2} \\
& \partial_+ \partial_- f_i=0.
\end{align}
\end{subequations}
The constraint equation is
\begin{equation}
e^{-2\phi /n} \bigg[ \frac{4}{n} \bigg(1-\frac{1}{n} \bigg) \partial_{\pm} \phi \partial_{\pm} \phi +\frac{2}{n} \partial_{\pm}^2 \phi -\frac{4}{n} \partial_{\pm} \rho \partial_{\pm} \phi \bigg] -\frac{1}{2} \sum_{i=0}^N \partial_{\pm} f_i \partial_{\pm} f_i=0.  \label{constraint}
\end{equation}
The expressions \eqref{variation1} and \eqref{variation2} implies that
\begin{equation}
\frac{2}{n} \partial_+ \partial_- (\rho-\phi)=0.  \label{eom2}
\end{equation}
Thus, we can still preserve the Kruskal gauge \eqref{kruskal gauge2}, for which $\rho=\phi$. The left expressions take the following form:
\begin{equation}
\partial_+ \partial_- (e^{-2\phi/n})=-1, \qquad \partial_{\pm}^2(e^{-2\phi/n})=-\frac{1}{2} \sum_{i=0}^N \partial_{\pm} f_i \partial_{\pm} f_i.  \label{eom3}
\end{equation}
At last, the general solution is obtained by (see \eqref{general solution1})
\begin{equation}
e^{-2\phi/n}=e^{-2\rho/n}=-x^+x^-+Q\log (-x^+x^-)+M.  \label{general solution2}
\end{equation}
In fact, the constant $M$ is related to the mass of the black hole, $M=\pi M_{\text{BH}}$. The constant $Q$ can be interpreted as the incoming and outgoing energy flux. If we substitute \eqref{constraint} to \eqref{eom2}, the stress tensor is given by $T_{\pm \pm} =\frac{Q}{(x^{\pm})^2}$.

\subsection{Static Black Holes}
\qquad For a static black hole with $Q=0$, which corresponds to an eternal black hole with unchanged temperature \footnote{The temperature of the FR model is special and always a constant \eqref{temperature}.}, or a black hole that is thermally at equilibrium with the thermal bath, the solution takes the following form:
\begin{subequations}
\begin{align}
e^{-2\phi/n}&=e^{-2\rho/n}=M-x^+x^-, \label{static} \\
\Omega_{\text{ete}}&=-x^+x^-+M-\frac{\kappa}{4} \log (-x^+x^-+M). \label{static conformal factor}
\end{align}
\end{subequations}
According to \eqref{critical}, to ensure that the singularity at $\Omega=\Omega_{\text{crit}}$ inside the apparent horizon, the mass $M$ is required to be $M>\frac{\kappa}{4}$. Besides, the event horizon at $x^+x^-=0$. Because the factor $\Omega$ can be viewed as the area of a black hole in 2D gravity, we can obtain the location of the apparent horizon by $\partial_+ \Omega=0$, which leads to\footnote{The other solution is $x^+x^-=M-\frac{\kappa}{4}$. However, under the limitation of a large mass, this location is near the singularity, which is unphysical and can be discarded.} $x^-=0$. Therefore, for an eternal black hole, the apparent horizon and the event horizon overlap at $x^+x^-=0$.

\subsection{Dynamical Black Holes}
\qquad Let us now return to the dynamical black hole that forms due to the collapse of a spherical shell of photons. Using the constraint equation \eqref{constraint}, we can obtain the general solution regarding some physical quantities, e.g., the Kruskal momentum and energy \cite{rst}
\begin{equation}
P_+(x^+)=\int_0^{x^+} dx^+ T_{++} (x^+),   \label{kruskal momentum}
\end{equation}
and
\begin{equation}
M(x^+)=\int_0^{x^+} dx^+ x^+ T_{++} (x^+).  \label{kruskal energy}
\end{equation}
Then we obtain
\begin{equation}
e^{-2\phi/n}=e^{-2\rho/n}=-x^+[x^-+P_+(x^+)]+M(x^+).
\end{equation}
The above equation provides a singularity at $M(x^+)-x^+(x^-+P_+(x^+))=0$. The event horizon is located at $x^-_H+P_+(\infty)=0$. Now, we consider an incoming shock wave with the stress tensor $T_{++}=\frac{1}{2} \sum_{i=0}^N \partial_+f_i \partial_- f_i=a \delta(x^+-x_0^+)$ at $x_0^+$. Then, geometries can be patched along this null trajectory. The corresponding solution is satisfied by
\begin{subequations}
\begin{align}
e^{-2\phi/n}&=e^{-2\rho/n}=-\frac{M}{x_0^+}\big(x^+-x_0^+\big) \theta \big(x^+-x_0^+\big) -x^+x^-, \label{dynamical} \\
\Omega_{\text{eva}}&=-x^+x^--\frac{M}{x_0^+} \big(x^+-x_0^+\big)  \theta \big(x^+-x_0^+\big) -\frac{\kappa}{4} \log (-x^+x^-), \label{dynamical conformal factor}
\end{align}
\end{subequations}
where $\theta$ is a step function and the parameter $a$ is given by $a=\frac{M}{x_0^+}$. Therefore, it is clear that below the region $x^+<x_0^+$, the geometry is vacuum \eqref{vacuum conformal factor}. In contrast, on the region $x^+>x_0^+$, the geometry corresponds to the static black hole \eqref{static conformal factor} discussed before.\\
\indent Similarly, this solution has an apparent horizon, which is defined by $\partial_+ \Omega=0$. Thus, the curve of the apparent horizon is given by\footnote{In the large mass limit, we can set $x_0^+=1$ to rescale the parameter $a$.}
\begin{equation}
(M-x^-)x^+=-\frac{\kappa}{4}.  \label{apparent horizon}
\end{equation}
We can calculate the coordinate of the endpoint of evaporation, which is the intersection of the critical curve \eqref{critical} and the curve of the apparent horizon \eqref{apparent horizon}, using the critical curve \eqref{critical} as the boundary of spacetime
\begin{equation}
(M+x^-_{\text{end}}) x^+_{\text{end}} =-\frac{\kappa}{4},
\end{equation}
and
\begin{equation}
-x^+_{\text{end}} x^-_{\text{end}}-M(x^+_{\text{end}}-1)-\frac{\kappa}{4} \log (-x^+_{\text{end}} x^-_{\text{end}})=\frac{\kappa}{4}-\frac{\kappa}{4} \log \frac{\kappa}{4}.
\end{equation}
At large mass, we obtain the coordinates of the endpoint
\begin{subequations}
\begin{align}
x^+_{\text{end}}&=\frac{(e^{\frac{4M}{\kappa}}-1)\kappa}{4M} \simeq \frac{\kappa e^{\frac{4M}{\kappa}}}{4M}, \label{endponit1} \\
x^-_{\text{end}}&=\frac{Me^{\frac{4M}{\kappa}}}{1-e^{\frac{4M}{\kappa}}} \simeq -M.  \label{event horizon}
\end{align}
\end{subequations}
The singularity becomes naked after the endpoint, which contradicts the cosmic censorship hypothesis. Spacetime as a whole is ill-defined. However, we ignore these considerations and only focus on the geometry before evaporation. Moreover, one should note that the curve \eqref{event horizon} also represents the event horizon because the apparent and event horizons are coincident at the endpoint.

\subsection{Coarse-grained and Fine-grained Entropy}
\qquad We briefly discuss two types of entropies and the related expressions\footnote{Strictly speaking, the fine-grained entropy at the finest level should be a constant in time in black hole without a bath. In this sense, the fine-grained entropy here is only more fine than the coarse-grained entropy. We thank Hao Geng for figuring out this point.}. The thermodynamic entropy or the Bekenstein--Hawking entropy is just the coarse-grained entropy. For 2D gravity, the Bekenstein--Hawking entropy can be derived by
\begin{equation}
S_{\text{BH}}=\frac{4\pi}{\sqrt{-g}} \frac{\partial \cal{L}}{\partial R}\Bigg |_{\text{horizon}}=e^{-2 \phi_H/n}=2M,  \label{bh entropy}
\end{equation}
where $\cal{L}$ is the Lagrangian of the action \eqref{fr action} and $\phi_H$ is denoted by the value of dilaton at the event horizon. Therefore, the Hawking temperature is given by
\begin{equation}
T_H=\frac{1}{2\pi}. \label{temperature}
\end{equation}
Now recall the island formula \eqref{island formula}, in which its complement is the fine-grained entropy of the \emph{black hole} based on the complementary of von-Neumann entropy. In this sense, the generalized entropy of a black hole is the Bekenstein-Hawking entropy plus the entropy contributed by matter fields surrounding the black hole's spacetime. For the matter field, the entanglement entropy, or equivalently, the von-Neumann entropy in vacuum CFT in flat spacetime $ds^2_{\text{flat}}=-dx^+dx^-$ is given by\footnote{Here CFT with the central charge $c=N$ is minimally coupled to gravity.}
\begin{equation}
S_{\text{matter}}=\frac{N}{3} \log [d(A,B)],
\end{equation}
where $d(A, B)$ is the proper distance between points $A$ and $B$ in flat spacetime, yielding
\begin{equation}
d(A,B)=\sqrt{[x^+(A)-x^+(B)][x^-(B)-x^-(A)]}.
\end{equation}
For the 2D metric $ds^2_{\ 2D}=-e^{2\rho(x^+,x^-)} dx^+dx^-$, the vacuum expectation value of normal ordering stress tensor can be written in terms of function $t_{\pm}$
\begin{equation}
\bra{\psi} : T_{\pm \pm} (x^{\pm}) : \ket{\psi} = -\frac{1}{12} t_{\pm} (x^{\pm}). \label{vev}
\end{equation}
Taking a conformal reparametrization for $x^{\pm} \to y^{\pm}$, the conformal factor transforms into
\begin{equation}
\rho (y^{\pm}) =\rho (x^{\pm})-\frac{1}{2} \log \frac{dy^+}{dx^+} \frac{dy^-}{dx^-},
\end{equation}
and the function $t_{\pm}$ transforms a Weyl rescaling
\begin{equation}
t_{\pm}(y^{\pm})= \bigg( \frac{dx^{\pm}}{dy^{\pm}} \bigg) t_{\pm} (x^{\pm}) +\frac{1}{2}  \{ x^{\pm}, y^{\pm} \},
\end{equation}
 with the Schwarzian derivative
 \begin{equation}
 \{ x^{\pm}, y^{\pm} \} \equiv \frac{x^{\prime \prime \prime}}{x^{\prime}} -\frac{3}{2}  \bigg( \frac{x^{\prime \prime}}{x^{\prime}} \bigg)^2.
 \end{equation}
Then, we can map the vacuum state from the flat metric $ds^2_{\text{flat}}$ to the curved metric $ds^2_{\ 2D}$ by the Weyl transformation. Thus, \eqref{vev} vanishes, then the function $t_{\pm}(x^{\pm})$ is zero. We further examine the behavior of von-Neumann entropy under this transformation \cite{bulk entropy}
 \begin{equation}
 S_{W^{-2}g} = S_g -\frac{N}{6} \sum_{\text{endpoints}} \log W,
 \end{equation}
where $W^{-2}=e^{-2\rho}$ and $g$ is the metric. Finally, we obtain the general expression for entanglement entropy of the single interval $[A, B]$ in 2D spacetime
 \begin{equation}
 S_{\text{matter}}=\frac{N}{6} \log \big[ d^2(A,B) e^{\rho(A)} e^{\rho (B)} \big] \Big |_{t_{\pm}=0}. \label{entanglement entropy}
 \end{equation}
Consequently, we use the above expression to calculate the entanglement entropy with and without islands in Sec.\ref{Evaporating Black Holes} and Sec.\ref{Eternal Black Holes}.

\section{Entanglement Entropy for Evaporating Black Holes} \label{Evaporating Black Holes}
\begin{figure}[htb]
\centering
\includegraphics[scale=0.3]{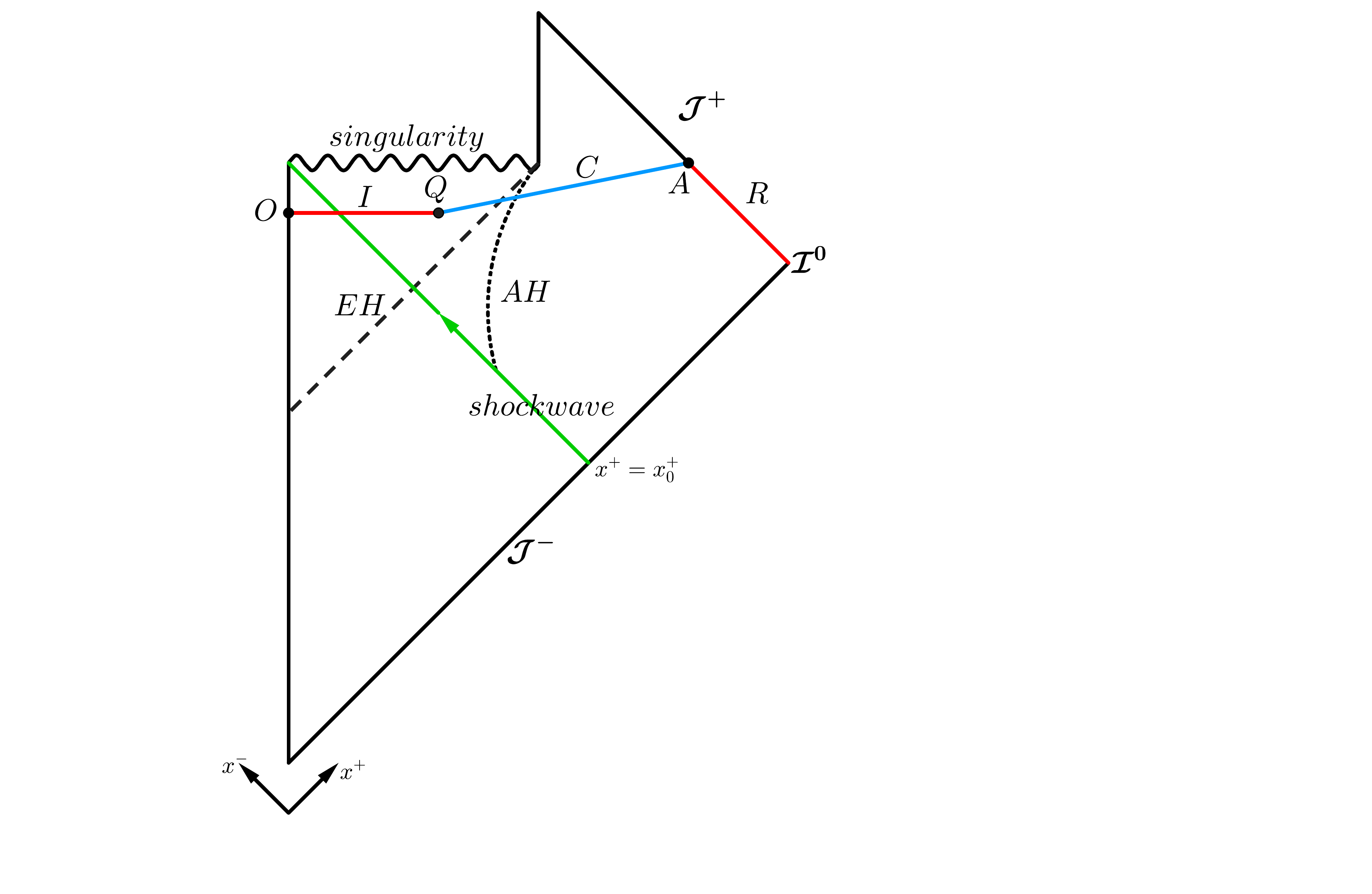}
\caption{\label{evap} The Penrose diagram of an evaporating black hole formed by null shell collapsing at $x^+=x_0^+(=1)$. ``EH'' and ``AH'' represent the event and apparent horizons, respectively. $\cal{I}^{\text{0}}$ and $\cal{J}^{\pm}$ are denoted as the space-like infinity and the future/past null infinity. The island is at point $Q$. The boundary of the radiation, i.e., the cutoff surface, is $A$. In the late time and the large distance limit, $A$ is approximately at $\cal{J}^+$, where the asymptotic observer collects the Hawking radiation. The region $C$ is the complementary of the union region $I \cup R$.}
\end{figure}

\qquad In this section, we use the island formula \eqref{island formula} to calculate the entanglement entropy of Hawking radiation for the evaporating black hole. We first consider the geometry without islands and then focus on the construction with an island. The Penrose diagram of evaporating black holes is shown in \mpref{evap}.

\subsection{Without island}
\begin{figure}[htb]
\centering
\includegraphics[scale=0.3]{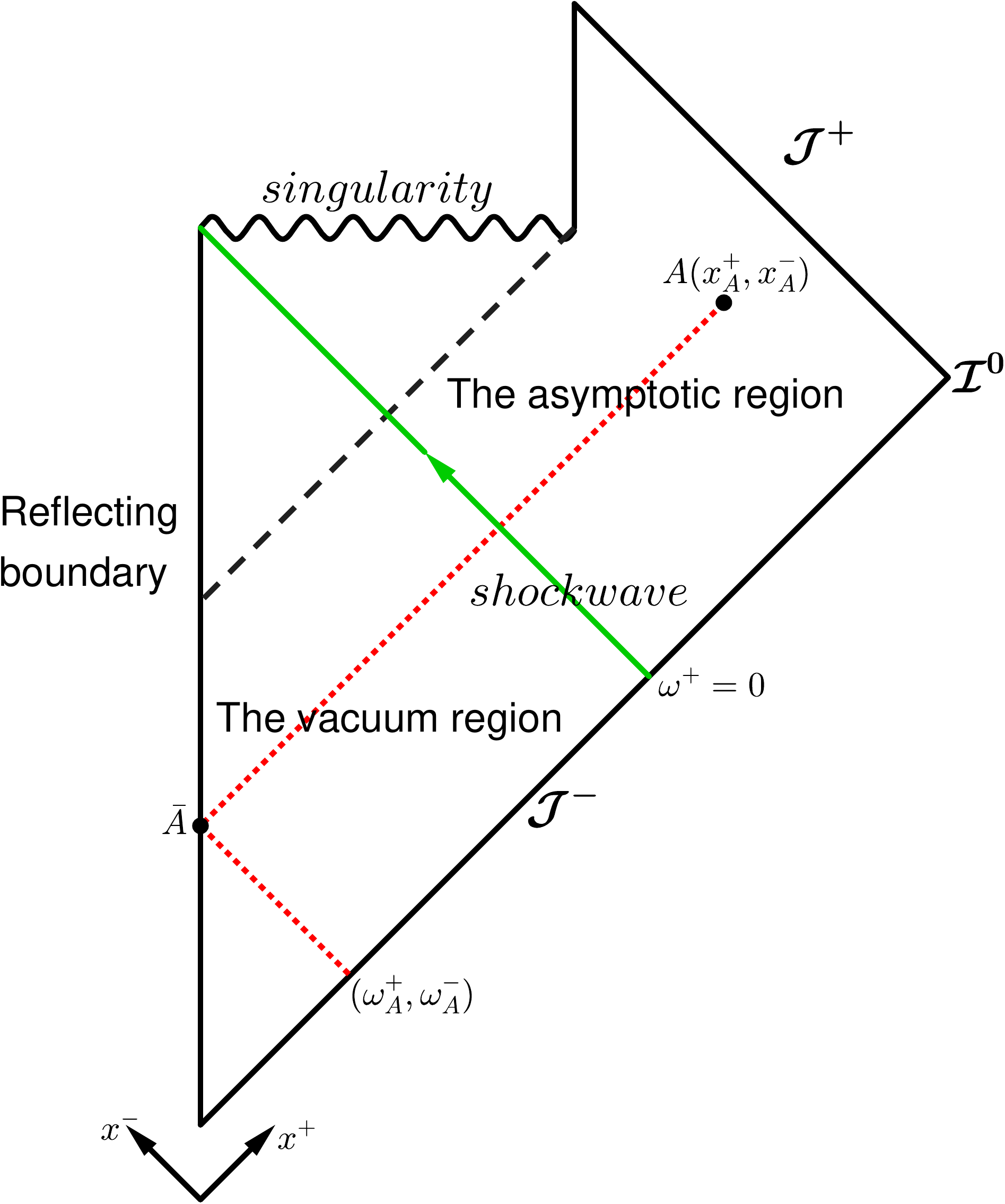}
\caption{\label{reflect} The diagrammatic sketch of the reflecting boundary. A point $A$ on the asymptotic region where $x^+>1$ is reflected to $\cal{J}^-$. Point $\bar{A}$ is the reflection point on the boundary. In such way, its coordinate is mapped to the ``ingoing'' frame \eqref{ingoing frame}. }
\end{figure}

\qquad In the absence of islands, the matter part of entanglement entropy $S_{\text{matter}}$ is only contributed by radiation. We assume that a black hole is formed by collapsing a pure shell of photons. Then, according to the complementary, we have $S_{\text{matter}}(R)=S_{\text{matter}}(C)$. At the boundary, we choose the reference point $O(x_o^+,x_o^-)$. The corresponding entropy $S_{\text{matter}}(C)$ can be calculated by \eqref{entanglement entropy}
\begin{equation}
S_{\text{matter}}(C) = \frac{N}{6} \log \big[ d_{\text{in}}^2(A,O) e^{\rho_{\text{in}}(A)} e^{\rho_{\text{in}}(O)} \big].  \label{eva no island entropy}
\end{equation}
The meaning of the subscript ``in'' will be explained later. For the convenience of calculation, we first introduce the following ``ingoing'' and ``outgoing'' coordinate frames
\begin{subequations}
\begin{align}
 &\text{near  $\cal{J}^-$}: \ \  x^+=+e^{+\omega^+}, \qquad x^-=-e^{-\omega^-},  \label{ingoing frame}  \\
 &\text{near  $\cal{J}^+$}: \ \  x^+=+e^{\sigma^+},  \qquad \ \ x^-=-e^{-\sigma^-}-M.  \label{outgoing frame}
\end{align}
\end{subequations}
In the region $x^+<1$, the vacuum solution \eqref{vacuum} in the frame $\{ \omega^{\pm} \}$ takes the form
\begin{equation}
ds^2=-(e^{\omega^+ - \omega^-})^{1-n} d \omega^+ d \omega^-.  \label{rindler}
\end{equation}
We can see that it corresponds to the Minkowski spacetime for $n=1$. However, for\footnote{We can take a coordinate transformation: $\omega^{\pm}=\eta \pm \xi$. Then the metric \eqref{rindler} becomes $ds^2=-e^{2(1-n)\xi} (d \eta^2 -d \xi^2)$, which is the standard Rindler metric for $n\ne 1$} $n\ne 1$, it corresponds to the Rindler spacetime.\\
\indent In the region $x^+>1$, the metric \eqref{dynamical} in the frame $\{ \sigma^{\pm} \}$ can be written as
\begin{equation}
ds^2=- \frac{e^{\sigma^+ - \sigma^-}}{(M+e^{\sigma^+ - \sigma^-})^n} d\sigma^+ d\sigma^-.
\end{equation}
Again, near the $\cal{J}^+$($\sigma^+ \to \infty$), we find that the above metric approaches the Rindler metric $ds^2(\sigma^+ \to \infty) \simeq -(e^{\sigma^+-\sigma^-})^{1-n} d\sigma^+ d\sigma^-$ unless $n=1$.\\
\indent Now, we follow the technique in \cite{2D information loss} that impose a reflecting condition at the time-like boundary (see the \mpref{reflect}). In this way, all region points $x^+>1$ are reflected to $\cal{J}^-$. Thus, the subscript ``in'' means that these quantities are only evaluated in the ``ingoing'' frame $\{ \omega^{\pm} \}$.
Therefore, we have
\begin{equation}
e^{2\rho_{\text{in}}(o)}=\bigg(e^{\omega^+(o)-\omega^-(o)} \bigg)^{1-n}= \big(-x^+_o x^-_o \big)^{1-n},
\end{equation}
\begin{equation}
e^{2\rho_{\text{in}}(A)}= \frac{-x^+_A x^-_A}{\big(M-x^+_A x^-_A -M x^+_A\big)^n},
\end{equation}
\begin{equation}
d^2(A,O)=\big( \omega^+(A) - \omega^+(O) \big) \big( \omega^-(O)-\omega^-(A) \big)= \log \frac{x_A^+}{x_O^+} \log \frac{x_A^-}{x_O^-}.
\end{equation}
Substituting these equations into the expression \eqref{eva no island entropy}, we obtain
\begin{equation}
\begin{split}
S_{\text{rad}} (\text{no island})&=\frac{N}{12}\log \Bigg[ \bigg( \log \frac{x_A^+}{x_O^+} \log \frac{x_A^-}{x_O^-} \bigg)^2  \big( -x_O^+ x_O^- \big)^{1-n}  \frac{-x_A^+ x_A^-}{\big(M-x_A^+x_A^--Mx_A^+\big)^n} \Bigg] \\
&\simeq \frac{N}{12} \log \bigg[ \frac{-x_A^+ x_A^-}{ \big(M-x_A^+ x_A^- -Mx_A^+ \big)^n}\bigg]\\
&=\frac{N}{12} \log \bigg(1+M e^{\sigma_A^-} \bigg) + \frac{N}{12} \log \bigg[ \big( \sigma_A^+ - \sigma_A^- \big)^{1-n} \bigg] \\
&\simeq \frac{N}{12} \sigma_A^- +(1-n) \frac{N}{12} \log \big(\sigma_A^+ \big). \label{eva result1}
\end{split}
\end{equation}
 As the cutoff surface is near $\cal{J}^+$ ($x_A^+ \to \infty$), all dependence on $x_O^{\pm}$ can be neglected. In addition, we rewrite the result in the ``outgoing'' coordinate $\sigma_A^-$ that represents the affine retarded time for the asymptotic observer. At the very early times, $\sigma_A \sim 0$, we have
\begin{equation}
S_{\text{rad}}(\text{very early times}) \simeq (1-n) \frac{N}{12} \log \big(\sigma_A^+ \big).  \label{early times1}
\end{equation}
Note that this is a divergent result for $n \ne 1$. But this divergent indeed can be absorbed in the UV cutoff term $\log \frac{1}{\epsilon_{\text{uv}}}$, where $\epsilon_{\text{uv}} \ll 1$ is the cutoff \cite{2d3}.\\
\indent However, at late times, as more and more Hawking pairs appear, this result is dominated by the first term, and the second $n$-dependent term is exponentially suppressed. Therefore, at late times, entropy increases linearly with the retarded time $S \propto \frac{N}{12}\sigma^-$. Then, the entropy without island violates the unitary principle. At late times, the fine-grained entropy should decrease. In particular, it can never be greater than the Bekenstein--Hawking entropy. Otherwise, the original black hole (after the Page time) does not have enough DOF to purify the thermal radiation. This is also called the ``bag of gold'' paradox \cite{review}.

\subsection{With Island}
\qquad We now introduce a modified semiclassical theory---the island paradigm---to reconcile the contradiction between the result \eqref{eva result1} and the fundamental principles of QM. We use the island formula \eqref{island formula} to recalculate the entanglement entropy of the radiation for the construction with an island and eventually reproduce a unitary Page curve.\\
\indent We still insist that the cutoff surface $A$ is near $\cal{J}^+$ and assume that the island $Q$ is located at the asymptotic region where $x^+>1$. For this construction, the union region $I\cup C \cup R$ (see \mpref{evap}) is considered the full Cauchy slice. According to the island formula \eqref{island formula}, the corresponding entropy of matter fields is determined by the union $I \cup R$, which is written as follows \cite{EE formula}
\begin{equation}
S_{\text{matter}} (R \cup I) = \frac{N}{6} \log \bigg[ e^{\rho_{\text{in}}(A)} e^{\rho_{\text{in}}(Q)}  \frac{d^2(A,Q) d(A,\bar{A})d(Q, \bar{Q})}{d(\bar{A},Q)d(A,\bar{Q})}  \bigg],
\end{equation}
where $\bar{A}$ and $\bar{Q}$ are the reflection point on the boundary. Although the expression looks complicated, we only focus on the behavior of the Page curve at late times. At late times and large distances, the following approximate relation exists
\begin{equation}
d(A,\bar{A}) \simeq d(Q,\bar{Q}) \simeq d(A,Q) \simeq d(\bar{A},\bar{Q}) \gg d(A,Q) \simeq d(\bar{A},\bar{Q}). \label{ope}
\end{equation}
Then, we obtain
\begin{equation}
\begin{split}
S_{\text{matter}} (R\cup I) &\simeq \frac{N}{6} \log \bigg[ e^{\rho_{\text{in}}(A)} e^{\rho_{\text{in}}(Q)} d^2(A,Q) \bigg] \\
&=\frac{N}{12} \log \Bigg[ \bigg( \log \frac{x_A^+}{x_Q^+} \log \frac{x_A^-}{x_Q^-} \bigg)^2  \frac{x_A^+x_A^-}{\big(M-x_A^+x_A^--Mx_A^+\big)^n}  \frac{x_Q^+x_Q^-}{\big(M-x_Q^+x_Q^--Mx_Q^+ \big)^n} \Bigg]. \label{eva island entropy}
\end{split}
\end{equation}
We then recall the Bekenstein--Hawking entropy $S_{\text{BH}}$ \eqref{bh entropy}, which corresponds to a quarter of the area of the black hole in 4D spacetime. Thus, we obtain the analogy of the gravity part
\begin{equation}
S_{\text{grav}}=\frac{\text{Area}(\partial I)}{4G_N}=2\bigg( \Omega_{\text{eva}}(Q)- \Omega_{\text{crit}} \bigg).  \label{eva grav entropy}
\end{equation}
Here, we introduce $\Omega_{\text{crit}}$ to ensure that the area of the island vanishes at the boundary. Correspondingly, the generalized entropy of the radiation is
\begin{equation}
\begin{split}
S_{\text{gen}}&=2\bigg( \Omega_{\text{eva}}(Q)- \Omega_{\text{crit}} \bigg)+ S_{\text{matter}}(R \cup I) \\
&=2M \bigg[ \bigg( 1- \frac{x_Q^+ x_Q^-}{M} -x_Q^+ \bigg) -\frac{\kappa}{4M} \log \big( -x_Q^+x_Q^- \big ) \bigg]\\
&+\frac{N}{12} \log \Bigg[ \bigg( \log \frac{x_A^+}{x_Q^+} \log \frac{x_A^-}{x_Q^-} \bigg)^2  \frac{x_A^+x_A^-}{\big(M-x_A^+x_A^--Mx_A^+\big)^n}  \frac{x_Q^+x_Q^-}{\big(M-x_Q^+x_Q^--Mx_Q^+ \big)^n} \Bigg].  \label{eva gen entropy}
\end{split}
\end{equation}
Extremizing the above expression with respect to $x_Q^+$ and $x_Q^-$ yields the following equations
\begin{subequations}
\begin{align}
\frac{\partial S_{\text{gen}}}{\partial x_Q^-} &= -2M \bigg( 1+\frac{x_Q^-}{M} \bigg) - \frac{N \big(M+(n-1)(M+x_Q^-)x_Q^+\big)}{12x_Q^+ \big(-M+(M+x_Q^-)x_Q^+\big)}-\frac{N}{6\log \bigg(\frac{x_A^+}{x_Q^+}\bigg)x_Q^+} =0,  \label{eva gen1}\\
\frac{\partial S_{\text{gen}}}{\partial x_Q^+} &=-2M x_Q^+ - \frac{N \big( M-(M+(1-n)x_Q^-)x_Q^+\big)}{12x_Q^- \big(-M+(M+x_Q^-) x_Q^+\big)}-\frac{N}{6\log \bigg( \frac{x_A^-}{x_Q^-} \bigg) x_Q^-}=0. \label{eva gen2}
\end{align}
\end{subequations}
As the point $A$ nears $\cal{J}^+$, i.e., $x_A^+ \to \infty$, the last term in \eqref{eva gen1} can be omitted. We also make the near event horizon limitation assumption to obtain the location of the island, where $x_Q^- \simeq -M$. Solving these two equations in the near event horizon limit, we obtain the following equation from \eqref{eva gen1}
\begin{equation}
x_Q^- \simeq -M+ \frac{N}{24x_Q^+}, \qquad x_Q^+ \simeq \frac{N}{24(M+x_Q^-)}.  \label{island location1}
\end{equation}
The large mass limit suppresses all $n$-dependent terms in the above solution. In this assumption, we also discover that the location of the island $x_Q^- \simeq -M+ \cal{O}$$\Big( \frac{1}{x_Q^+} \Big)$ is indeed very close to the event horizon and inside the event horizon (see \eqref{event horizon}). Then, substituting the solution \eqref{island location1} into \eqref{eva gen2}, we obtain
\begin{equation}
\log \bigg( \frac{x_A^-}{x_Q^-} \bigg) = \frac{2(M+x_Q^-)}{M}, \qquad x_A^-=-M-\frac{N}{24 x_Q^+}.  \label{island location2}
\end{equation}
On the other side, compared \eqref{island location2} with the ``outgoing'' coordinate $x_A^-=-e^{\sigma_A^-}-M$ \eqref{outgoing frame}, the following approximations are obtained as
\begin{equation}
\sigma_A^- =\log \frac{24x_Q^+}{N} \sim \frac{M}{N},  \qquad x_Q^+ \sim N e^{\frac{M}{N}}.  \label{island location3}
\end{equation}
The first approximation is made possible because the lifetime of black holes and asymptotic observer's retarded time $\sigma_A^-$ are of the same order. From \eqref{event horizon}, the lifetime is
\begin{equation}
\sigma_A^- \sim \sigma_{\text{end}}^-= -\log (-x_{\text{end}}^--M)=\frac{4M}{\kappa}-\log M \simeq \frac{48M}{N} \sim \frac{M}{N}.  \label{lifetime}
\end{equation}
Finally, the generalized entropy after the extremization is given by \eqref{eva gen entropy}, \eqref{island location1},\eqref{island location2}, and \eqref{island location3}
\begin{equation}
\begin{split}
S_{\text{gen}} &\simeq 2M -\frac{N}{12} -\frac{N}{24} \log \bigg( NM e^{\frac{M}{N}}-\frac{N}{24} \bigg) +{\cal O} \bigg( \log \big(M^{2-2n}\big) \bigg),  \\
&\simeq 2M-\frac{N}{24} \sigma_A^- = S_{\text{BH}}- \frac{N}{24} \sigma_A^-,
\end{split}
\end{equation}
where the parameter $n$ is also exponentially suppressed at the large mass and late times limit. Therefore, we obtain the entanglement entropy with the island as a function of the retarded time $\sigma^-$. Combining the result \eqref{eva result1}, we have
\begin{equation}
\begin{split}
S_{\text{rad}}&=\text{Min} \big[ S_{\text{gen}} (\text{no island}), S_{\text{gen}} (\text{island}) \big] \\
&=\text{Min} \bigg( \frac{N}{12} \sigma^- ,2M-\frac{N}{24} \sigma^- \bigg), \label{eva result2}
\end{split}
\end{equation}
which leads to a unitary Page curve.\\
\indent In conclusion, we reproduce the Page curve of an evaporating black hole using the island paradigm. At early times, the QES is a trivial surface lying on the boundary, which results in an increase in entanglement entropy curve that is linearly proportional to the retarded time \eqref{eva result1}. However, the island appears inside the event horizon after the Page time. The QES now is a nontrivial surface, leading to a transition in the entropy. At late times, the entanglement entropy is approximately equal to the Bekenstein--Hawking entropy and drops to zero when the black hole is evaporated. This result is also consistent with the Page curve derived from the Page theorem \cite{Page theorem}.

\section{Entanglement Entropy for Eternal Black Holes} \label{Eternal Black Holes}
\qquad In this section, we repeat the calculation procedure in the Sec.\ref{Evaporating Black Holes} to calculate the Page curve for an eternal black hole. The information paradox for eternal black holes is a deformed version of the black hole information paradox. The corresponding solution is more strict: at late times, a black hole in the Hartle--Hawking state has infinite entanglement with the outside thermal Hawking radiation, which far exceeds the entropy bound of a black hole \cite{Entropybound}. Similarly, we first calculate the entanglement entropy without the island and then consider the island.

\subsection{Without Island}
\begin{figure}[htb]
\centering
\includegraphics[scale=0.25]{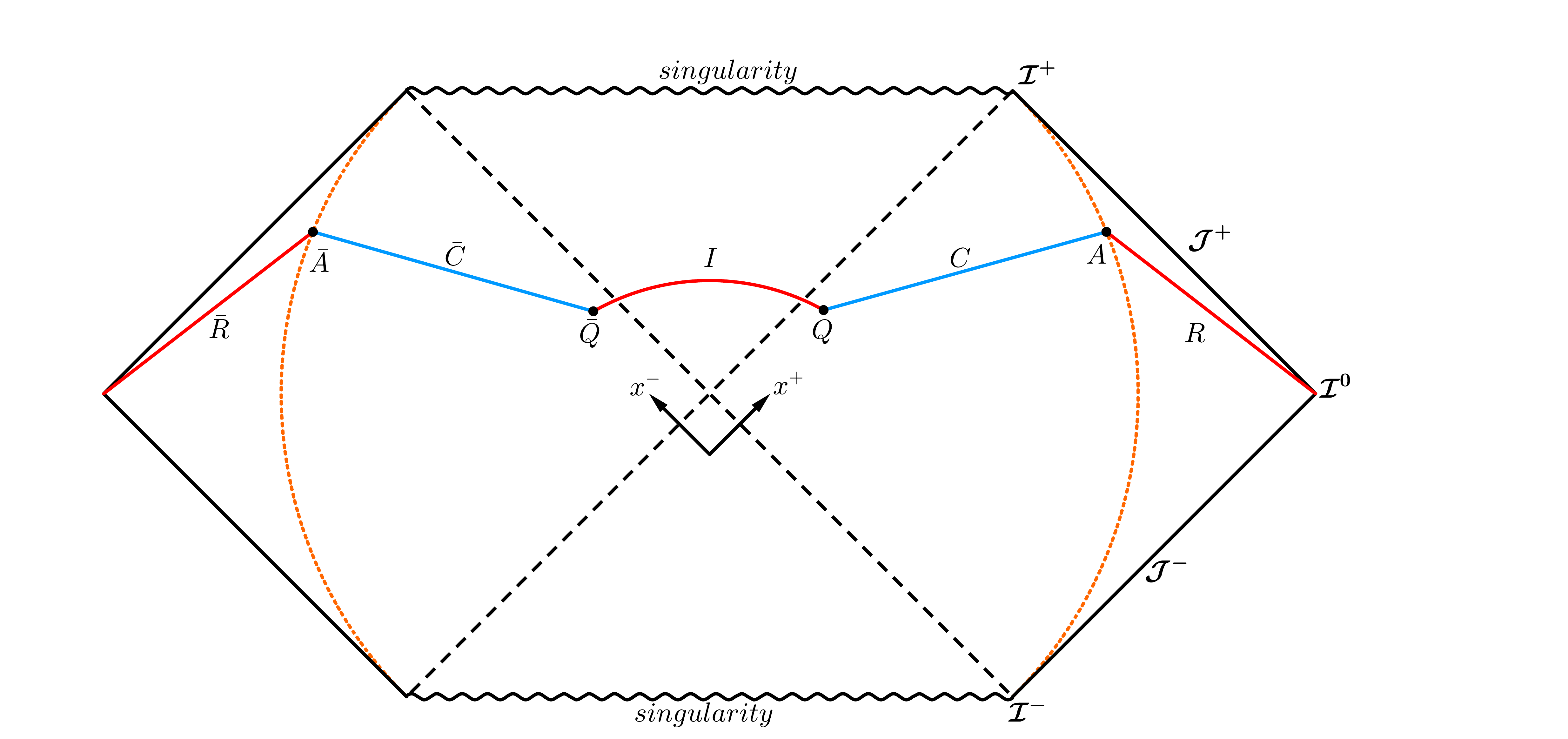}
\caption{\label{ete} The Penrose diagram for eternal black holes. The black dashed line represents the event horizon, and the orange dashed line represents the cutoff surface. The boundaries of the island and the radiation are denoted as $Q$ and $I$, respectively. Their symmetry points are denoted as $\bar{Q}$ and $\bar{A}$. The regions $C$ and $\bar{C}$ are complementary.}
\end{figure}

\qquad The Penrose diagram for the eternal black hole is shown in the \mpref{ete}. We only calculate the entanglement entropy of the union $R \cup \bar{R}$ for the no-island construction. According to the complementary of von-Neumann entropy, the entropy for the matter part is
\begin{equation}
S_{\text{matter}}(R)=S_{\text{matter}} (R \cup \bar{R})=\frac{N}{6} \log \bigg[ e^{\rho(A)} e^{\rho(\bar{A})} d^2 (A, \bar{A})  \bigg].  \label{ete no island entropy}
\end{equation}
Then, we transform to the asymptotically flat frame and make the following coordinate transformation:
\begin{subequations}
\begin{align}
&\text{The right wedge}: \ \ \ \  x^+=+e^{y+t}, \qquad x^-=-e^{y-t}, \label{right wedge} \\
&\text{The left wedge}: \qquad  x^+=-e^{y-t}, \qquad x^-=+e^{y+t}, \label{left wedge},
\end{align}
\end{subequations}
where $t$ and $y$ are time and spatial coordinates for the asymptotic observer near $\cal{J}^+$, respectively. For the eternal solution, we obtain the following expression from \eqref{static}
\begin{equation}
\begin{split}
S_{\text{rad}}(\text{no island})&=S_{\text{matter}} (R \cup \bar{R}) \\
&=\frac{N}{12} \log \Bigg[ \frac{\big(x_A^+ -x_{\bar{A}}^+ \big)^2 \big(x_A^--x_{\bar{A}}^- \big)^2 }{\big(M-x_A^+x_A^- \big)^n \big( M-x_{\bar{A}}^+ x_{\bar{A}}^-\big)^n} \Bigg]\\
&=\frac{N}{6} \log \big(4 \cosh^2 t_A \big) +\frac{N}{6} \log \Bigg[ \frac{e^{2y_A}}{\big(e^{2y_A}+M\big)^{\frac{1+n}{2}}} \Bigg]
\end{split}
\end{equation}
Similarly, at very early times, this result has a dependency on the parameter $n$, which behaves as the following
\begin{equation}
S_{\text{rad}} (\text{very early times}) \simeq \frac{N}{6}t_A^2 +\frac{N}{6} (3+n) y_A \sim  (3+n) y_A.  \label{early times2}
\end{equation}
However, at late times and large distances limit, the expression becomes
\begin{equation}
S_{\text{rad}}(\text{no island}) \simeq \frac{N}{3} t_A + \frac{N}{3} y_A- \frac{N}{12}(1+n) \log \bigg(e^{2y_A}+M \bigg) \simeq \frac{N}{3} t_A.  \label{ete result1}
\end{equation}
Again, the contribution from the parameter $n$ is a subleading term and is exponentially suppressed. Thus, the entropy without the island grows linearly with time $S\propto \frac{N}{3}t$ at late times. Then, the information paradox is sharpened here, the entanglement entropy never stops increasing and becomes divergent at late times.

\subsection{With Island}
\qquad We still expect the island paradigm to protect the bound of the entanglement entropy at late times. For the configuration with an island, the generalized entropy of the radiation is twice the area term plus the von-Neumann entropy of the CFT, which is
\begin{equation}
S_{\text{gen}}=2S_{\text{grav}}(Q)+S_{\text{matter}} (I \cup R).  \label{ete island entropy}
\end{equation}
At late times and large distances, we still preserve the approximation \eqref{ope}, which allows us to evaluate the entropy of the matter sector only in the interval $[\bar{A},\bar{Q}] \cup [A, Q]$. Furthermore, due to the symmetry, we only need to calculate one of the intervals. Thus, we have
\begin{equation}
\begin{split}
S_{\text{gen}} &\simeq 4 \bigg( \Omega_{\text{ete}}(Q) - \Omega_{\text{crit}} \bigg) + 2S_{\text{matter}}(C) \\
&=4(-x_Q^+ x_Q^-+M) +\frac{N}{6}  \log \frac{\big(x_A^+-x_Q^+\big)^2 \big(x_A^--x_Q^-\big)^2}{\big(M-x_A^+x_A^-\big)^n \big(M-x_Q^+ x_Q^-\big)^n}.  \label{ete gen entropy}
\end{split}
\end{equation}
We extremize the expression \eqref{ete gen entropy} with respect to $x_Q^{\pm}$, which gives
\begin{subequations}
\begin{align}
\frac{\partial_{\text{gen}}}{\partial x_Q^+} &=-4x_Q^- -\frac{N}{3\big(x_A^+ -x_Q^+ \big)} +\frac{Nnx_Q^-}{6\big(M-x_Q^+x_Q^- \big)}=0,  \label{ete gen1} \\
\frac{S_{\text{gen}}}{\partial x_Q^-} &=-4x_Q^+-\frac{N}{3\big(x_A^--x_Q^-\big)} + \frac{Nnx_Q^+}{6 \big(M-x_Q^+x_Q^-\big)}=0. \label{ete gen2}
\end{align}
\end{subequations}
Then, putting $A$ to $\cal{J}^+$ ($x_A^+ \to \infty$), we obtain
\begin{equation}
\frac{x_Q^+}{x_Q^-}=\frac{x_A^+}{x_A^-}, \qquad x_Q^{\pm} \simeq \frac{-N}{12x_A^{\mp}}, \label{island location4}
\end{equation},
which implies that the island is near and outside the horizon as $x_Q^+ \cdot x_Q^- \lesssim 0$. Finally, the entropy with the island is obtained by substituting \eqref{island location4} to \eqref{ete gen entropy}
\begin{equation}
\begin{split}
S_{\text{gen}} &\simeq 4M - \frac{N^2}{36 x_A^+ x_A^-} +\frac{N}{6} \log \frac{\big(-x_A^+ x_A^- \big)^{2-n}}{M^n} \\
&\simeq 4M + \frac{N}{3} y (2-n) -\frac{nN}{6} \log M \\
&\simeq 2S_{\text{BH}}. \label{ete result2}
\end{split}
\end{equation}
Clearly, at late times, the emergence of islands curbs the growth of the entanglement entropy. Therefore, we reproduce the Page curve for eternal black holes from the results \eqref{ete result1} and \eqref{ete result2}. We also find that the eternal case is very similar to the evaporating case. The only difference is that the entanglement entropy with an island for eternal black holes reaches a saturation value at late times and does not change with time. This is because the coupled auxiliary bath or incident boundary conditions continuously replenish the energy to the black hole. In addition, it is surprising that the location of the island is outside the event horizon. However, this result is consistent with the quantum focusing conjecture \cite{qfc}. Accordingly, there is a quantum teleportation protocol by which we can extract the information from the island, and one can refer to \cite{eternal bh, teleportation} for details.

\section{Page Curve and Scrambling Time} \label{Scrambling Time}
\qquad In this section, we plot the Page curve and derive the scrambling time. Then, we provide some remarks on the Page curve for 2D gravity.

\subsection{Evaporating Case}
\qquad For evaporating black holes, the lifetime is defined by \eqref{lifetime}, which is $\sigma_{\text{life}} \simeq\frac{4M}{\kappa}=\frac{48M}{N}$. The fine-grained entropy of the radiation is
\begin{equation}
\begin{split}
S_{\text{rad}}&=\text{Min} \bigg[ \frac{N}{12} \sigma^-, 2M-\frac{N}{24} \sigma^- \bigg] \\
&=\frac{N}{24} \text{Min} \big[ 2\sigma^-, \sigma_{\text{life}}-\sigma^- \big]. \label{fine entropy1}
\end{split}
\end{equation}
We obtain the Page time as
\begin{equation}
t_{\text{Page}}^{\text{eva}}=\frac{1}{3} \sigma_{\text{life}}=\frac{16M}{N}.  \label{pagetime1}
\end{equation}
The corresponding Page curve is shown in \mpref{evapage}. \\
\indent We now calculate the scrambling time. According to the Hayden--Preskill experiment, if Alice throws a quantum diary into the black hole after the Page time, she must wait for the so-called scrambling time to recover this information from the Hawking radiation \cite{HP}. As suggested by the entanglement wedge reconstruction, the scrambling time corresponds to the time when the infalling information hits the boundary of the island \cite{entanglement wedge}.\\
\indent After the Page time, the observer on the cutoff surface sends a light signal to the black hole at $x^+=x_L^+$. We assume that the cutoff surface has the same order as the event horizon, which is\footnote{We define the region inside the cutoff surface as the ``near black hole'' region. In the initial Hayden--Preskill experiment, this region is about 2 or 3 Schwarzschild radius for 4D Schwarzschild black holes \cite{HP}.}
\begin{equation}
\begin{split}
\Omega_{\text{near}}=(1+\alpha) \Omega_H &\simeq (1+\alpha) \big[ -x_H^+x_H^- -Mx_H^+ +M\big] \\
&=(1+\alpha)M = \frac{1}{2} (1+\alpha) S_{\text{BH}},  \label{near region},
\end{split}
\end{equation}
where $\alpha$ is a constant with the order ${\cal O}(1)$. $\Omega_H$ is the conformal factor at the event horizon, $x_H=-M$. Correspondingly, the infalling information at the cutoff surface satisfies
\begin{equation}
-x_L^+(x_L^-+M)+\frac{1}{2}S_{\text{BH}} =\frac{1}{2} (1+\alpha) S_{\text{BH}}.
\end{equation}
In the ``outgoing'' coordinate \eqref{outgoing frame}, the initial time is
\begin{equation}
\sigma_L^- \equiv \log \bigg( \frac{1}{-x_L^--M} \bigg) =\log \bigg( \frac{2x_L^+}{\alpha S_{\text{BH}}} \bigg).
\end{equation}
When the information reaches the island, the corresponding retarded time is
\begin{equation}
\sigma_Q^- \equiv \log \bigg( \frac{1}{-x_Q^--M} \bigg) =\log \bigg( \frac{24x_Q^+}{N} \bigg) =\log \bigg( \frac{24x_L^+}{N} \bigg).
\end{equation}
Then, the scrambling time is
\begin{equation}
\begin{split}
t_{\text{scr}}&=\sigma_Q^- -\sigma_L^- = \log \bigg( \frac{12\alpha S_{\text{Bh}}}{N} \bigg) \\
&\simeq \log \frac{S_{\text{BH}}}{N} \sim \frac{\beta}{2\pi} \log S_{\text{BH}}. \label{scr time1}
\end{split}
\end{equation}
Here, $\beta$ is the inverse temperature $\beta=\frac{1}{T_H}=2\pi$ \eqref{temperature}. This result is consistent with \cite{scrambling time, acoustic}

\subsection{Eternal Case}
\begin{figure}[htb]
\centering
\subfigure[\scriptsize{}]{\label{evapage}
\includegraphics[scale=0.55]{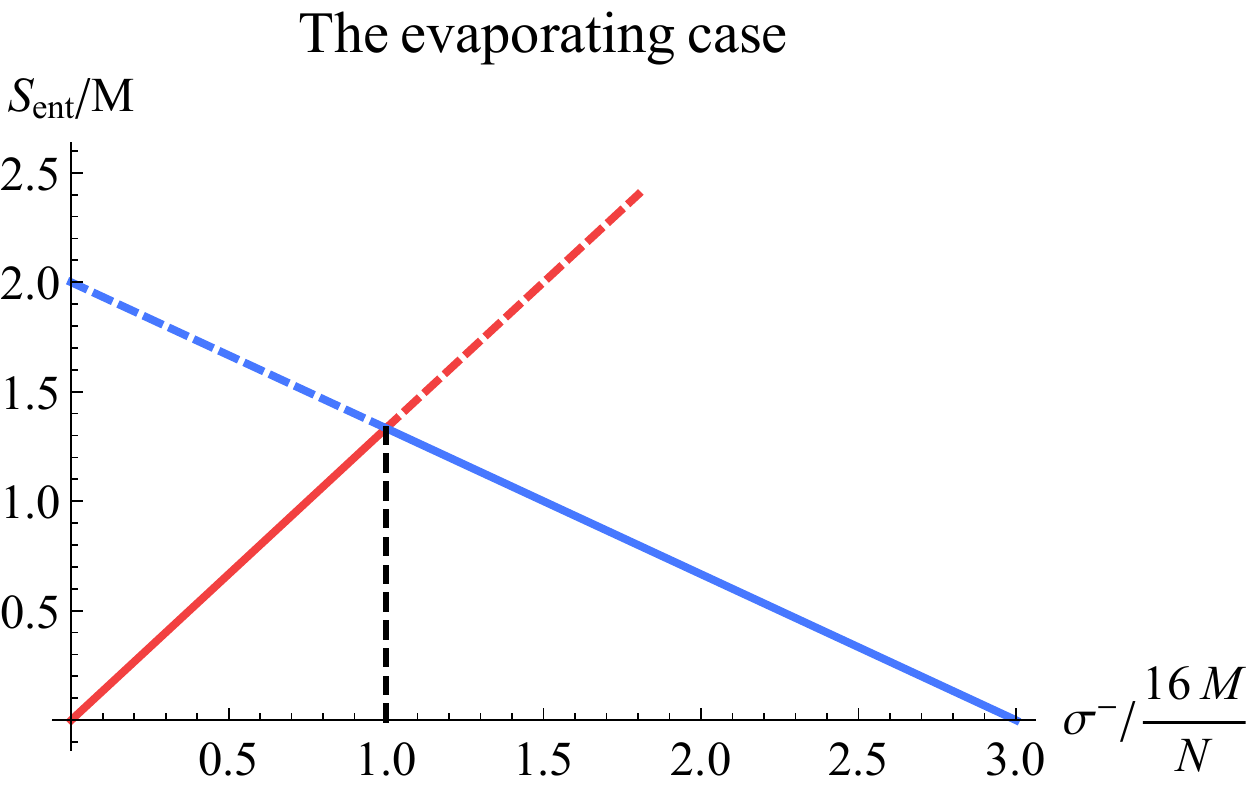}
}
\quad
\subfigure[\scriptsize{}]{\label{etepage}
\includegraphics[scale=0.55]{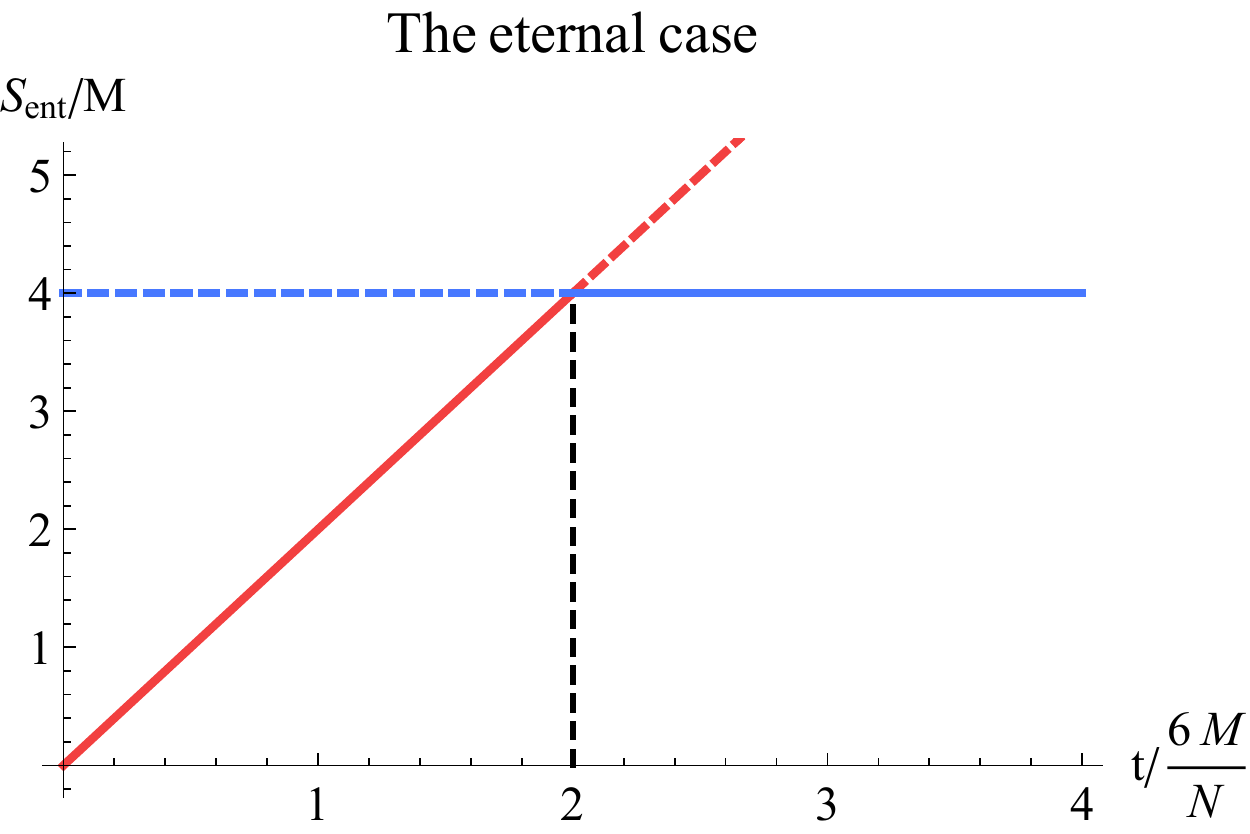}
}
\caption{Page curves for black holes. Red lines represent the entanglement entropy without islands, while blue lines represent the entropy by considering the island. On the left, the Page time for 2D evaporating black holes is approximately one-third of the lifetime. On the right, the entropy for eternal black holes grows to twice the Bekenstein--Hawking entropy after the Page time.}
\label{Page curves}
\end{figure}

\qquad The calculation for the eternal case is simpler. Comparing \eqref{ete result1} with \eqref{ete result2}, we obtain the fine-grained entropy of the radiation in the whole process as
\begin{equation}
S_{\text{rad}} = \text{Min}  \bigg[ \frac{N}{3}t, 4M \bigg].  \label{fine entropy2}
\end{equation}
So the Page time is
\begin{equation}
t_{\text{Page}}^{\text{ete}}= \frac{12M}{N}. \label{pagetime2}
\end{equation}
Compared with \eqref{pagetime1}, we can still consider that it has the same order of lifetime, even though the lifetime of an eternal black hole is indeed infinite. The corresponding Page curve is shown in \mpref{etepage}. \\
\indent Similarly, if we emit a light signal from the cutoff surface at time $t_0$, the signal approaches the island at time $t_1$. The scrambling time is determined by
\begin{equation}
\begin{split}
t_{\text{scr}}=t_1-t_0 &\simeq \log \bigg( \frac{x_Q^+}{x_L^+} \bigg) =\log \bigg( \frac{\alpha S_{\text{BH}}}{6N} \bigg) \\
& \sim \frac{\beta}{2\pi} \log S_{\text{BH}},  \label{scr time2}
\end{split}
\end{equation}
where $\alpha$ has the same meaning as in \eqref{near region}, and $x^+$ is the asymptotic coordinate \eqref{right wedge}. Here, we assume the spatial coordinate of the cutoff surface is approximately equal to the island, namely $y_L \simeq y_Q$.\\
\indent Let us now make some remarks. Different parameter $n$ for this generalized 2D model corresponds to different theories. However, at late times, we can find that the effect of $n$ is neglectable as the subleading term in both evaporating \eqref{fine entropy1} and eternal cases \eqref{fine entropy2}. This also suggests that the Page curve for the FR model is similar to the RST model \cite{2d2,2d3} at the leading order. The fundamental reason is that the Hawking temperature \eqref{temperature} is the same in both models because the original Page curve is derived from the Page theorem with the unitary assumption \cite{Page theorem}: At early times, the dimensions of the Hilbert space of the radiation were small. However, the dimensionality of the black hole was very big. The entanglement entropy of the radiation is approximate to the thermodynamic entropy, which is proportional to the Hawking temperature $T_H$. At late times, the radiation dominates, hence, the radiation and the black hole exchange the status. The entanglement entropy is finally approximated to the Bekenstein--Hawking entropy determined by the temperature $T_H$. We can provide a simplified version of the proof by the second law of thermodynamics. The total entropy change during the whole evaporation process is
\begin{equation}
\begin{split}
\Delta S_{\text{tot}}&=\Delta S_{\text{BH}} +\Delta S_{\text{rad}} \\
&=\int \frac{dM_{\text{BH}}}{T_H}=\int \frac{dM}{\pi T_H} = 2M.
\end{split}
\end{equation}
We fix the constant of integration by requesting that $S_{\text{BH}}=0$ as $M_{\text{BH}}=0$. After the evaporation, the black hole entropy changes to $\Delta S_{\text{BH}}=-S_{\text{BH}}=-2M$. Then, we obtain
\begin{equation}
\frac{\Delta S_{\text{rad}}}{\Delta S_{\text{BH}}}= \frac{\Delta S_{\text{tot}}-\Delta S_{\text{BH}}}{|\Delta S_{\text{BH}}|}=2.
\end{equation}
The entropy of the radiation increases twice as fast as the decreases of the black hole entropy, which leads to the Page time being one-third of the lifetime. This result is consistent with our calculation by the island paradigm. However, one should note that in the final evaporation stage, the semiclassical approximation is invalid. The quantum gravity effects are expected to play a major role. Therefore, the shape of the last part of the Page curve for evaporating black holes may not be suitable.

\section{Revisit the Firewalls and State Paradox} \label{Firewalls}
\qquad Thus far, we have reproduced the Page curve for evaporating and eternal black holes through the island formula \eqref{island formula}. The most important point is that when we calculate the von-Neumann entropy from matter fields, the contribution from the island region inside the black hole\footnote{Note that we use the cutoff surface to divide the regions inside and outside black holes. Therefore, although the island is outside the event horizon for eternal black holes, it is inside the region of black holes.} is considered. Equivalently, the entanglement wedge of the island is now a part of the entanglement wedge of the radiation. In some sense, this seems to derive a theory about the interior of the black hole. The interior structure of black holes remains mysterious in modern physics. The nonescaping information properties of the black hole interior forbid our understanding of the black hole interior. Now, the island inside the black hole may shed light on the firewall paradox \cite{firewall}, which was proposed a few years ago. Therefore, in this section, we first review the firewalls paradox and then discuss its relationship with the island paradigm. We show that the island can be a better candidate for solving the firewall paradox. Finally, we discuss the state paradox that the island formula may still cause. \\
\indent The firewall paradox is a variant version of the information paradox proposed by Almheiri, Marolf, Polchinski, and Sully (AMPS). The emergence of the firewall further intensifies the contradiction between the classical GR and QM near the event horizon. On the one hand, classical GR does not expect that there exist such strange objects like firewalls near the event horizon. On the other hand, the principle of quantum entanglement monogamy has to be maintained by introducing a firewall \cite{monogamy}.\\
\begin{figure}[htb]
\centering
\includegraphics[scale=0.3]{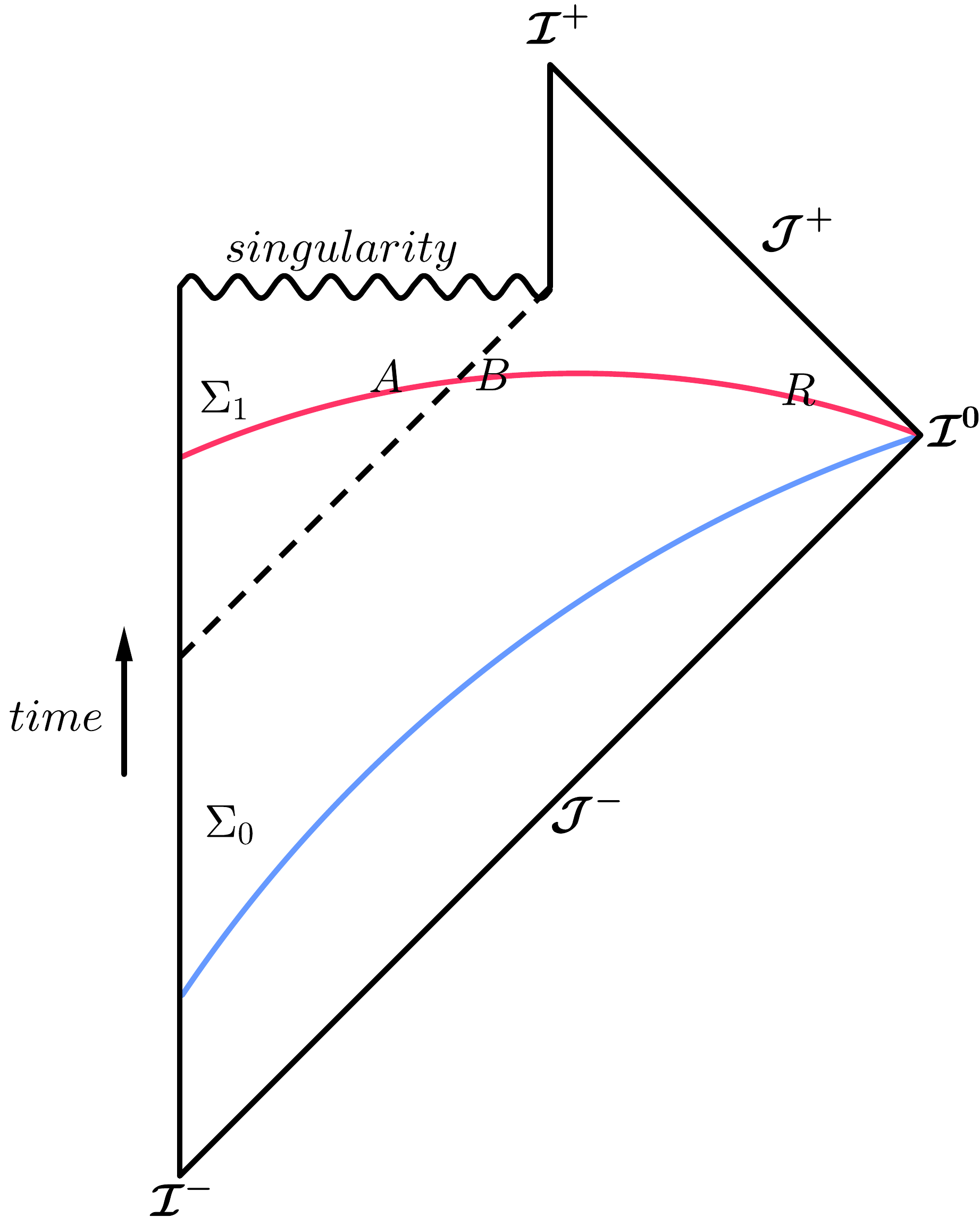}
\caption{\label{firewalls} The schematic diagram of the firewall. $\Sigma_0$ and $\Sigma_1$ are the Cauchy slices at early and late times, respectively. We divide $\Sigma_1$ into three parts, where $A$ represents the ``ingoing'' Hawking mode inside the horizon, i.e., the black hole region or the non-emitted Hawking radiation, $B$ represents the late Hawking radiation, and $R$ is the early Hawking radiation or the region of the radiation.}
\end{figure}
\indent Let us now review the firewall paradox in detail. Consider an old black hole after the Page time. We choose a full Cauchy slice $\Sigma_1$ and divide it into three regions, $A$, $B$, and $R$, as shown in \mpref{firewalls}.\\ We assume that the black hole is formed by pure quantum states. If the unitarity is respected, the late Cauchy slice $\Sigma_1$ evolves from the early Cauchy $\Sigma_0$. Therefore, the Hawking radiation must also be in the pure state as the unitarity also represents the map from one pure state to another pure state \cite{cauchy slice}. Then, the entanglement entropy is given by
\begin{equation}
S(A \cup B \cup R) \equiv S_{ABR}=0, \qquad S(B\cup R) \equiv S_{BR}=0.
\end{equation}
From \mpref{firewalls}, the combined region $A\cup B\cup R$ is the Cauchy slice, which has zero entanglement entropy, because the Cauchy slice contains all information in spacetime. The second relation is based on the above discussion that the Hawking radiation should be in a pure state. From the strong subadditivity inequality
\begin{equation}
\begin{split}
S_{ABR} \le S_A+S_{BR}=S_A, \qquad S_{ABR} \ge |S_A-S_{BR}| =S_A.   \label{inequality}
\end{split}
\end{equation}
We finally obtain $S_{ABR}=0=S_A$. The mutual information between the $A$ and $BR$ is\footnote{When we are not sure whether subsystems are pure or not, a better candidate for measuring entanglement is the mutual information.}
\begin{equation}
I(A;BR) \equiv S_A +S_{BR}-S_{ABR}=0,
\end{equation}
which means that the subsystem $A$ has no correlation with the subsystem $BR$ or the reduced density matrix $\rho_{ABR} = \rho_A \otimes \rho_{BR}$. However, according to the unitary evaporation, the just-emitted Hawking radiation $B$ is highly entangled with the mode $A$ inside the black hole. Thus, the mutual information should satisfy $I(A;BR) \gg 1$. Then, there is a clear contradiction between the two statements. If there is no entanglement between the inside and outside black holes, $A$ and $BR$ are disconnected. The wavefunction of the radiation $\psi$ near the event horizon behaves as
\begin{equation}
\partial_x \psi \big|_{x_H} \sim \frac{1}{\epsilon},
\end{equation}
where $x_H$ is the location of the event horizon and $\epsilon$ is the cutoff. Then, the Hamiltonian of wavefunction has the following form
\begin{equation}
{\cal H} \sim \int \big( \partial_x \psi \big)^2 dxdy \sim \frac{A}{\epsilon^3} \to \infty.
\end{equation}
Therefore, the energy at the event horizon is very high. AMPS claims that the event horizon is like a firewall, which cuts off the entanglement between $A$ and $BR$. Simultaneously, the appearance of the firewall also implies that the smoothness of horizons is destroyed, which is also not allowed by the classical GR.\\
\indent However, even if the firewall cannot exist, we inevitably encounter other paradoxes. If the evaporation is not a unitarity process, the principle of QM is also violated. Because there are no firewalls, $A$ and $B$ are highly entangled. Simultaneously, the early time Hawking radiation $R$ is maximally entangled with the late time Hawking radiation $B$ from the Page theorem. This violates the principle of entanglement monogamy. \\
\indent In brief, the firewall leads to conflict between GR and QM. Although many different solutions are proposed, all have some flaws \cite{quantum computation, apologia}. Further, we revisit the issue from the island paradigm point of view. Based on the entanglement wedge reconstruction \cite{entanglement wedge}, the \emph{interior} of black holes belongs to the \emph{outside} radiation. Thus
\begin{equation}
A \in R, \qquad S_{ABR} = S_{BR}.
\end{equation}
We substitute this equation back into \eqref{inequality} and obtain
\begin{equation}
\begin{split}
S_{ABR}&=S_{BR} \le S_A+S_{BR}, \\
S_{ABR}&=S_{BR} \ge |S_A-S_{BR}|, \\
\text{Cauchy slice}: \ \ S_{ABR}&=S_{BR}=0.
\end{split}
\end{equation}
Then, we have
\begin{equation}
S_A=S_{BR}=0.
\end{equation}
Therefore, the outgoing Hawking radiation, including the interior, is in a pure state, which is consistent with the unitarity. Besides, as the interior is a part of the radiation, the entanglement between the late Hawking radiation and the early Hawking radiation amounts to the entanglement between the late Hawking radiation and the interior, which is not contrary to the entanglement monogamy and does not lead to the firewall. Thus, it also ensures the smoothness of the horizon.\\
\indent Hence, the island seems to be a better candidate than firewalls. Moreover, we should note that whether islands or firewalls, they are singular objects at the horizon, and they exist to avoid the occurrence of the paradox. Therefore, we only reserve one of them but not both in the description of the evolution of black holes \cite{QI}. However, even though the island remains, one still encounters some issues. Back to the island formula \eqref{island formula}, we rewrite it by the entanglement wedge proposal
\begin{equation}
S_{\text{rad}}=\frac{\text{Area}(\partial \text{EW})}{4G_N}+S [\text{EW} (\text{rad})],   \label{island formula ew}
\end{equation}
where ``EW'' is denoted as the entanglement wedge. We can see that ``rad'' exists on both sides of the equation. The radiation is in the \emph{full} quantum description on the left. However, on the right, the radiation is still in the \emph{semiclassical} frame, i.e., Hawking's view (a mixed state). In particular, we still take Hawking's initial state, which leads to the Hawking curve as the input when we calculate the fine-grained entropy of the radiation. By contrast, we only contain the island in the entanglement wedge of the radiation after the Page time, which lead to the Page curve (a pure state). As the $S$-matrix that describes the evolution of black holes is an observable quantity, the corresponding state of the Hawking radiation should not be ambiguous. Such a conflict is called the state paradox \cite{state paradox}.\\
\indent A solution to the conflict is to introduce the gravity/ensemble duality \cite{ensemble}. We assume that the theory of gravity corresponds to a series of field theories. Then, the single pure state evolves into different pure states under every Hamiltonian. The linear combination of these pure states can constitute a mixed state. This mixed state corresponds to Hawking's initial input. As a result, once we introduce the ensemble, we need to redefine the entropy. The way is that one first performs an ensemble average for all pure states and obtains a mixed state $\braket{\rho}$. If we calculate the entanglement entropy $S(\braket{\rho})$ for this mixed state, the result corresponds to the Hawking curve, the same as Hawking's calculation. However, we first evaluate the entanglement entropy $S(\rho_i)$ corresponding to each pure state and then implement the ensemble average for this entropy to obtain $\braket{S(\rho)}$. This result is consistent with unitarity, as each entropy represents a unitary evolution. Therefore, the later result is self-averaging, which reproduces the Page curve. Accordingly, the island formula \eqref{island formula ew} can be modified into the following form
\begin{equation}
\braket{S[\rho(t)]} = \frac{\text{Area} \big[  \partial \text{EW} \big( \braket{\rho(t)}  \big) \big]  }{4G_N}+S \big( \text{EW} (\braket{\rho(t)}) \big). \label{modify island formula}
\end{equation}
More specific content is beyond the scope of this paper, one can refer to \cite{ensemble} for more information.

\section{Discussion and Conclusion} \label{Conclusion}
\qquad In this paper, we investigated the information paradox for the generalized 2D gravity. Unlike the famous RST model, the FR model represents a family of exactly solvable theories, and the corresponding geometries are also interesting. We calculate the entanglement entropy of the Hawking radiation by using the island paradigm in this generalized 2D background. No matter whether it is the evaporating black hole or the eternal black hole, the island is always absent at early times, which results in the entropy of the radiation growing proportional to the time (\eqref{eva result1} and \eqref{ete result1}). The behavior of the entanglement entropy satisfies the Hawking curve. However, a Page transition occurs at the Page time when the entropy increases to the Bekenstein--Hawking entropy bound. In our calculations, the minimality condition requires us to attribute the island region inside the black hole to the entanglement wedge of the outside radiation. After the Page time, the area term is dominant. The island is close to and inside the event horizon for the evaporating black hole. The entanglement entropy drops to zero eventually \eqref{eva result2}. The corresponding Page time is about a third of the lifetime of black holes (\mpref{evapage}). The island is close to and outside the event horizon for the eternal black hole in the Hartle--Hawking state. The entanglement entropy finally reaches a saturation value of twice the Bekenstein--Hawking entropy \eqref{ete result2}. The Page curve of this case is shown in \mpref{etepage}. However, the impact of the parameter $n$ in the FR model is significant only at very early times (\eqref{early times1} and \eqref{early times2}). Once we consider the late times and the large distance limits, the contribution of the parameter is only a subleading term. It is exponentially suppressed in both the evaporating and the eternal case. This result leads to Page curves being the same as the RST model in the leading order. The essence is that the Hawking temperature of the two models is the same and does not depend on the parameter $n$. The generalized second law can also verify this, and the result is consistent with the original derivation of Page curves through the Page theorem. Therefore, the calculations in this paper are meaningful and extend the scope of the application of the island paradigm.\\
\indent Besides, we also discuss the connection between the firewall and the island. The price of firewalls is the collapse of the effective field theory or the classical GR near the horizon. Although this is still an open question, we show that islands appear to be a better candidate to replace the firewall from the perspective of the entanglement wedge reconstruction. At last, we briefly review the state paradox. The island provides a DOF that can purify the thermal radiation, which somewhat solves the information paradox. However, the island formula does not modify any calculations of the original Hawking's calculation but introduces a non-local DOF such as islands in the black hole interior. Therefore, there are still some issues called the state paradox. The gravity/ensemble duality is an attempt to solve the issue. By introducing the ensemble average to redefine the entropy, we finally obtain the modified island formula \eqref{modify island formula}. \\
\indent At last, some of the following issues require our attention in the future:\\
1. The aspect of the calculation. First, we only consider the large mass limit for evaporating black holes, hence, the $n$-dependent term is discarded. But at the end of evaporation, the $n$-dependent term is no longer the subleading term and significantly impacts the result. In particular, the semiclassical approximation is breakdown when the mass approaches the Planck scale. The anti-Page curve may appear on the Planck scale \cite{anti pc}. The physical details at this point are expected to be dominated by quantum gravity, which we do not yet know. Second, although the island formula is derived from the AdS black hole with couples the auxiliary bath, its applicability extends far beyond AdS spacetime. However, baths are not introduced in our model. The black hole in the FR model is asymptotically flat. Therefore, the Hawking radiation naturally propagates to the null infinity. Nonetheless, we insist on the importance of baths. The associated calculations may fail without baths \cite{2d5}. Third, so far, most related work that studies the island of evaporating black holes focus on 2D gravity. On the one hand, the calculation corresponding to higher-dimensional black holes is difficult, especially as the entanglement entropy formula is divergent in high dimensions. On the other hand, there are no known analytical solutions for higher-dimensional asymptotically flat evaporating black holes. We expect some related work in the future, such as the classical Vaidya
 spacetime. Fourth, the island formula can be obtained by the gravitational path integral, in which a new saddle point is needed to consider, called the replica wormholes saddle. We can also investigate the dynamic evolution of the replica wormhole geometry in the Sachdev--Ye--Kitaev (SYK) model based on the JT/SYK dual. There are also some interesting reports \cite{syk1,syk2,syk3}.\\
2. The aspect of quantum information. According to ``$ER=EPR$'' \cite{EPR}, the entanglement between the DOF of two subsystems produces a connected geometry--wormholes that bridges them to each other. However, the island formula only provides the evolution of Page curves. Still, it does not explain how the quantum information escapes from the black hole to the Hawking radiation. We can only treat the island formula as a black box operation.\\
3. The aspect of the ensemble theory. Some study demonstrates that the ensemble that corresponds to the gravity comes from the baby universe inside the black hole \cite{baby1,baby2,baby3,baby4,baby5,baby6}. However, this is still debated. The core issue is understanding the dynamics inside the black hole. However, there is currently no effective means to detect these dynamics.\\
\indent In conclusion, the current theories describing the evaporation process of black holes are imperfect, especially in explaining what happens when a black hole is near the end of evaporation. Perhaps the discovery of quantum gravity or some other new physical mechanism in the future can explain this conundrum.

\section*{Acknowledgement}
We would like to thank Haiming Yuan for discussions related to quantum field theories, Cheng Ran for discussions related to the ensemble theory, and Wencong Gan for discussions related to the Vaidya spacetime. We would also like to thank Hao Geng for useful comments. The study was partially supported by NSFC, China (grant No.11875184).

\begin{appendix}
\section{Brief review of the FR model}  \label{appendix}
\qquad In this appendix, we display more details about the FR model and briefly discuss the geometry for the special case of $n=0$ and $n=2$. \\
\indent The metric of static black hole spacetime can be read from \eqref{static}.
\begin{equation}
ds^2=-e^{2\rho}dx^+dx^-=-\frac{1}{\big(M-x^+x^- \big)^n} dx^+dx^-.  \label{metric A}
\end{equation}
The Ricci curvature scalar $R$ is
\begin{equation}
R=8e^{-2\rho}\partial_+ \partial_- \rho =4Mn \big( M-x^+x^- \big)^{n-2}.  \label{Ricci}
\end{equation}
It is convenient to recast the metric \eqref{metric A} in the chiral form by the Eiddton transformation.
\begin{equation}
x^+=V, \qquad  x^+x^-=r,
\end{equation}
which is followed by
\begin{equation}
V=e^v, \qquad x=\frac{\big( M-r\big)^{1-n}}{2(1-n)}=\frac{M-x^+x^-}{2(1-n)},
\end{equation}
which yields the chiral form
\begin{equation}
\begin{split}
ds^2&= \bigg \{ 2(1-n)x-M\bigg[2(1-n)x \bigg]^{\frac{n}{n-1}} \bigg \} dv^2 +2dv dx \\
&=-f(x)dv^2+2dvdx,  \label{chiral metric}
\end{split}
\end{equation}
where $f(x)$ is the metric function, whose derivative is
\begin{equation}
\begin{split}
f^{\prime}(x)&=2(1-n) \bigg \{ 1-M \bigg( \frac{n}{n-1} \bigg)  \bigg[ 2(1-n)x \bigg] \bigg \}^{\frac{1}{1-n}}, \\
f^{\prime \prime}(x)&=-4Mn \bigg[ 2(1-n)x \bigg]^{\frac{2-n}{n-1}}.    \label{derivative}
\end{split}
\end{equation}
Then, the Ricci curvature scalar $R$ is simply in the chiral coordinates, $R=-f^{\prime \prime}(x)=4Mn \big[ 2(1-n)x \big]^{\frac{2-n}{n-1}}=4Mn \big(M-x^+x^- \big)^{n-2}$, and the dilaton $\phi$ gives the coupling.
\begin{equation}
e^{2\phi /n}= \big[ 2(1-n)x \big]^{\frac{1}{n-1}}=\frac{1}{M-x^+x^-}.  \label{coupling}
\end{equation}
Now, we rewrite the metric in the Schwarzschild gauge $d^2=-f(x)dt^2+f^{-1}(x)dr^2$. Then, the event horizon can be obtained by $f(x_H)=0$, which obtains
\begin{subequations}
\begin{align}
x_H &= \frac{\big( M-x_H^+x_H^- \big)^{1-n}}{2(1-n)}=\bigg(\frac{1}{M}\bigg)^{n-1} \frac{1}{2(1-n)}, \\
x_H^+x_H^-&=0.  \label{event horizon A}
\end{align}
\end{subequations}
Thus, the Hawking temperature is given by
\begin{equation}
T_H=\frac{f^{\prime}(x_H)}{4\pi}=\frac{1}{2\pi}.   \label{temperature A}
\end{equation}
The singularity is located as
\begin{equation}
x_s^+x_s^-=M, \label{singularity}
\end{equation}
for $0<n<2$.
For $n=0$, we find that the spacetime is flat according to \eqref{Ricci}. However, it is not a trivial result because the coupling \eqref{coupling} is nontrivial and becomes singular. One should rescale the dilaton $\phi$ to $ \tilde{\phi}=n \phi$ first and then take the limit $n \to 0$. Accordingly, the classical action \eqref{fr action} takes the following form:
\begin{equation}
\tilde{S}_0 = \frac{1}{2\pi} \int d^2 x \sqrt{-g}  \bigg( e^{-2 \tilde{\phi}} R +4 \lambda^2 \bigg). \label{cghs}
\end{equation}
If we redefine the metric tensor by $g_{\mu \nu} \to e^{2 \phi} g_{\mu \nu}$, this action becomes the CGHS action \cite{cghs}. Therefore, the special case $n=0$ represents an unusual black hole with a space-like singularity in the coupling but the Ricci curvature vanishes. For the concrete content of this case, one can refer to \cite{fr1}. For the other case $n=2$, the curvature scalar is a constant \eqref{Ricci}. However, a similar interpretation for the $n=0$ case as applies here. The dilaton is singular again. However, in the following discussion, we see that the curvature of the $n=2$ case is no longer constant when considering the quantum effects.\\
\indent Next, we return to the dynamical black hole with the back-reaction. The back-reaction effects are well explained in the RST model \cite{rst}. Nevertheless, some nontrivial quantities still have parameter-dependent evolution in the FR model. Therefore, we give some discussions of the differences.\\
\indent The usual time-dependent geometry that describes the collapse of a massless shock wave and then evaporates is in terms of $\Omega$.
\begin{equation}
\Omega=\chi=-x^+[x^- +P_+ (x^+)] -\frac{\kappa}{4} \log \big( -x^+x^- \big) +M(x^+).
\end{equation}
Moreover, the Ricci curvature is
\begin{equation}
R=4ne^{-2\rho} \frac{1}{e^{-2 \phi/n} -\frac{\kappa}{4}}  \bigg( 1-\frac{4}{n^2} \partial_+ \phi \partial_- \phi e^{-2\phi/n} \bigg).
\end{equation}
We find there exists a singularity at the following line
\begin{equation}
\phi = \phi_{\text{crit}}=-\frac{n}{2} \log \frac{\kappa}{4}.    \label{critical phi}
\end{equation}
When $T_{++}<\frac{\kappa}{4} (x^+)^2$, the line is time-like. However, it turns to space-like if $T_{++}>\frac{\kappa}{4} (x^+)^2$. Substituting \eqref{critical phi} to the conformal factor $\Omega$ by \eqref{kruskal gauge2} and \eqref{conformal factor2}, we obtain \eqref{critical} eventually. We should note that in the case $n=2$, it represents the spacetime with a constant curvature at the classical level. However, once the one-loop correction is considered, the curvature is divergent at $\phi_{\text{crit}}$ rather than a constant because the coupling becomes strong. Besides, the metric also deviates from the classical theory in the strong-coupling region. For more information, one can refer to \cite{fr1,fr2}, and we end our discussion here.

\end{appendix}

\end{document}